\documentclass{sigchi}

\toappear{Permission to make digital or hard copies of all or part of this work for personal or classroom use is granted without fee provided that copies are not made or distributed for profit or commercial advantage and that copies bear this notice and the full citation on the first page. Copyrights for components of this work owned by others than ACM must be honored. Abstracting with credit is permitted. To copy otherwise, or republish, to post on servers or to redistribute to lists, requires prior specific permission and/or a fee. Request permissions from Permissions@acm.org.\\
{\emph{CSCW '16}}, February 27-March 02, 2016, San Francisco, CA, USA. \\
\copyright 2016 ACM. ISBN 978-1-4503-3592-8/16/02\$15.00 \\
DOI: http://dx.doi.org/10.1145/2818048.2820019}

\usepackage{balance}  
\usepackage{graphics} 
\usepackage{txfonts}
\usepackage{times}    
\usepackage[pdftex]{hyperref}
\usepackage{color}
\usepackage{textcomp}
\usepackage{booktabs}
\usepackage{todonotes}
\usepackage{multirow}
\usepackage{subfigure}
\usepackage[font=small,labelfont=bf]{caption}
\makeatletter
\def\url@leostyle{%
  \@ifundefined{selectfont}{\def\UrlFont{\sf}}{\def\UrlFont{\small\bf\ttfamily}}}
\makeatother
\def\pprw{8.5in}
\def\pprh{11in}

\definecolor{linkColor}{RGB}{6,125,233}
\hypersetup{%
  pdftitle={SIGCHI Conference Proceedings Format},
  pdfauthor={LaTeX},
  pdfkeywords={SIGCHI, proceedings, archival format},
  bookmarksnumbered,
  pdfstartview={FitH},
  colorlinks,
  citecolor=black,
  filecolor=black,
  linkcolor=black,
  urlcolor=linkColor,
  breaklinks=true,
}

\begin{document}

\title{\#Bieber + \#Blast = \#BieberBlast: \\Early Prediction of Popular Hashtag Compounds}

\numberofauthors{3}
\author{
  \alignauthor{Suman Kalyan Maity\\
    \affaddr{Dept. of CSE}\\
    \affaddr{IIT Kharagpur, India}}\\
  \alignauthor{Ritvik Saraf\\
    \affaddr{Dept. of Maths \& Computing}\\
    \affaddr{IIT Guwahati, India}}\\
  \alignauthor{Animesh Mukherjee\\
    \affaddr{Dept. of CSE}\\
    \affaddr{IIT Kharagpur, India}}\\
}

\maketitle

\begin{abstract}
Compounding of natural language units is a very common phenomena. In this paper, we show, for the first time, that Twitter hashtags which, could be considered as correlates of such linguistic units, undergo compounding.  We identify reasons for this compounding and propose a prediction model that can identify with 77.07\% accuracy if a pair of hashtags compounding in the near future (i.e., 2 months after compounding) shall become
popular. At longer times $T$ = 6, 10 months the accuracies are 77.52\% and 79.13\% respectively. This technique has strong implications to trending hashtag recommendation since newly formed hashtag compounds can be recommended early, even before the compounding has taken place. Further, humans can predict compounds with an overall accuracy of only 48.7\% (treated as baseline). Notably, while humans can discriminate the relatively easier cases, the automatic framework is successful in classifying the relatively harder cases.
\end{abstract}



\section{Introduction}
Hashtag is the new ``paralanguage'' of Twitter. What started as a way for people to connect with others and to
organize similar tweets together, propagate ideas, promote specific people or topics has now grown into a language of its own. As hashtags are created by people on their own, any new event or topic can be referred to by a variety of hashtags. This linguistic innovation in the form of hashtags is a very special feature of Twitter which has become immensely popular and are also widely adopted in various other social media like Facebook, Google+ etc. and have been studied extensively by researchers to analyze the competition dynamics, the adoption rate and popularity scores. However, there are very few attempts to study the linguistic aspects of hashtag evolution over large time scales.
One of the interesting and prevalent linguistic phenomena in today's world of brief expressions, chats etc. is hashtag compounding where new hashtags are formed through combination of two or more hashtags together with the form of the individual hashtags remaining intact. For example, \#PeoplesChoice and \#Awards together form \#PeoplesChoiceAwards. \#KellyRipa and \#CelebrationMonth make \#KellyRipaCelebrationMonth; \#WikipediaBlackout is formed from \#Wikipedia and \#Blackout; \#OregonBelieveMovieMeetup is formed from \#Oregon, \#BelieveMovie and \#Meetup; \#Educational, \#Ipad, \#Apps together make \#EducationalIpadApps etc. In this paper, we identify for the first time that while some of these compounds gain a high frequency of usage over time (even higher than the individual constituents) many of them are soon lost into oblivion. We focus and investigate in detail the reasons behind the above observations.
\subsection*{Motivations}
In etymology, we come across a very similar phenomenon where words are formed from various other words sampled from the same or a different language. Lexical compounding has been prevalent all through over the history of evolution of any language~\cite{bad,compound,gaeta}. For example, in English, `wheelchair' has been formed from `wheel' and `chair', bookworm is the combination of `book' and `worm' with the meaning of the words getting completely modified due to compounding. Similarly, `in so far' has become `insofar' with no meaning getting altered. However, such compounding phenomena in social media are far more prevalent than in standard texts and language. 

Innovation and adoption are both important processes in language change~\cite{Croft2000,Milroy1992}. While innovation refers to the creation of new linguistic units, adoption refers to its proliferation among wider groups of speakers. An innovative form must be adopted by a significant number of speakers in order for
observable change to take place~\cite{Milroy1992}. Hashtag is a linguistic innovation in social media. Predicting the propagation and spread of hashtags in online communities is an important aspect from both commercial and psychological perspectives. Not all hashtags become popular, some of them become popular while most of them fall into oblivion. There are numerous factors that drive hashtag popularity and this popularity aspect have been studied extensively by various researchers~\cite{kamath,kong,bigbird,sun,maon,what,Lilian2012srep,filippo1,yang}. A significant proportion of the hashtags used in social media are compound hashtags. In this paper, we attempt to early predict whether a hashtag compound shall gain a higher usage frequency (popularity) than the individual constituent hashtags forming the compound. Note that our objective does not include identifying if two or more hashtags are going to compound in future; instead, we are interested to automatically identify cases where the popularity of an 
already formed compound is far more than the individual components.

\subsubsection*{Compounds in Practice}
Like general hashtags, predicting popular hashtag compounding is also an important and interesting task. There are marketing strategic needs, needs for fulfilling communicative intents (affective expression, political persuasion, humor etc.) as well as spontaneous needs for use of hashtag compounds. For example, the e-commerce company Amazon used \#AmazonPrimeDay to promote the discounted sale of its product. The hashtag is a compound of \#Amazon and \#PrimeDay whereas the individual hashtag \#PrimeDay was also popular. So, there is a trade-off whether to use hashtag compounds or the uncompounded constituents. Similarly, assume another scenario where an event is taking place, say the premiere of a movie `The Imitation Game'. Here one can use both the hashtags \#TheImitationGame and \#Premiere or can use a hashtag compound \#TheImitationGamePremiere. In this context, one needs to identify which version one should use so that the hashtag being used gains a higher frequency of usage in the near future. \#CSCW2016 is being used to tag the activities taking place related to the 2016 CSCW conference. This is also 
a compound hashtag made of \#CSCW and \#2016 where \#CSCW refers to all CSCW conferences and \#2016 refers to all the events/activities going to take place in 2016. The hashtag \#CSCW2016 is used for a more focused purpose and refering to only the 2016 edition of the conference whereas \#CSCW could also have served the purpose. Hashtag compounds also serve the communicative intents like political campaign hashtags (\#PresidentTrump = \#President + \#Trump : hashtag that shows support for Donald Trump for the 2016 US Presidential election). Hashtag compounding also happen spontaneously. These hashtags are generally conversational or personal themed hashtags like \#TheBestFeelingInARelationship (\#TheBestFeeling + \#InARelationship), \#ThrowbackThursday (\#Throwback + \#Thursday), \#ComeOnNowDontLie (\#ComeOnNow + \#DontLie).

Our prediction framework is different from existing popularity prediction/trend identification algorithms/frameworks in the following ways. The popularity prediction frameworks deal with the problem of predicting whether a hashtag will become popular or not among a competing hashtag pool consisting of all hashtags across various topics from the data stream and filtered by the time window in which the prediction is being made. However, in our framework, we want to predict whether the hashtag compound or the individual constituent hashtags become popular. Therefore, our competition space is smaller and topically more well-defined. This is simply to say whether to adopt the compounded hashtags or not.

\subsection*{Research objectives and contributions}
In this paper, we study the hashtag compounding phenomena as a linguistic innovation and investigate in detail the socio-linguistic reasons for its adoption. Towards this objective, we make the following contributions in the paper: 
\begin{itemize}
 \item We study the hashtag compounding phenomena for the first time and put forward various socio-linguistic reasons for the adoption (popularity) of the compound hashtags
 \item We conduct a thorough experiment with human subjects to identify how well humans can predict popular hashtag compounds; the accuracy obtained is 48.7\% and constitutes the baseline.
 \item We finally use the socio-linguistic aspects as features in a model that is able to predict popular future hashtag compounds ($T$ = 2 months) with an overall accuracy of 77.07\% which is $\sim$58\% improvement over the baseline. Note that our results have the potential to strongly impact the trending hashtag recommendation application of Twitter since it is able to predict hashtags (i.e., compounds) that will be
popular in future even before the hashtags are born (i.e., before the compounding has taken place).
\item We also perform long term predictions at $T$ = 6 and 10 months after compounding and achieve 77.5\% and 79.13\% accuracy respectively.
\item Finally, we perform a thorough correspondence analysis of the prediction outcomes from human evaluations and the automatic framework. We observe striking differences between the outcomes; while human evaluators are usually able to discriminate the relatively easier cases, the automatic framework is very successful in distinguishing some of the harder cases. We argue that this is a methodological novelty of this paper and can be adopted in future experimental studies of similar type.
\end{itemize}

\subsection*{Organization of the paper}
The remainder of the paper is organized as follows. Section 2 is a concise review of the state-of-the-art. In section 3, we describe the dataset briefly. In section 4, we discuss about the adoption/popularity of hashtag compounds. Section 5 investigates the different linguistic aspects responsible for hashtag compounding. In section 6, we outline the baseline experiments based on the human judgments. In section 7, we introduce the prediction model and describe the features. In section 8, we evaluate the model and discuss the discriminative power of the features. In section 9, we perform a detailed correspondence analysis of the hashtags judged by human evaluators and by the automatic prediction framework. In section 10, we discuss the implications of the findings from our study. Finally, in section 11, we conclude and point to future direction of research.

\section{Related work}
\subsection*{Language use in social media}
There have been considerable works that focus on the content and its linguistic aspects in social media. Honeycutt and Herring~\cite{hohe} analyzed conversational exchanges in Twitter focusing on mentions. Ritter et al.~\cite{alan} developed an unsupervised learning approach to identify conversational structure from open-topic conversations. Danescu-Niculescu-Mizil et al.~\cite{dan} studied how people adopt linguistic styles while in conversation on Twitter. Eisenstein et al.~\cite{eisen} studied the role of geography and demographics on the language in Twitter. Hong et al.~\cite{hong} investigated the cultural differences in Twitter's language. Hu et al.~\cite{hu} studied the characterization of linguistic and psycholinguistic aspects in Twitter. Wang et al.~\cite{wenbo} studied how people curse each other in Twitter. Almuhimedi et al.~\cite{alm} performed a large scale quantitative analysis on deleted tweets.

There have been several studies on how language is used in social communities. Kramer et al.~\cite{kramer} characterized different types of discourse in online support groups (specifically, emotion writing, talkative, bipolar chat) for successful communities. Arguello et al.~\cite{argue} assessed whether members were likely to post again using linguistic features of their posts. Nguyen et al.~\cite{nguyen} proposed a novel approach to identify latent hyper-groups in social communities based on users' language use. Cassell and Tverky~\cite{cassell} described how linguistic interaction patterns change over time. Matthews et al.~\cite{matt} characterized how online communities combine multiple social tools. Tausczik and Pennebaker~\cite{penn} study the motivation of people participation in Q\&A sites (MathOverflow) and found that building reputation is an important incentive. Matthews et al.~\cite{hews} studied the relationship between member satisfaction and language use within content posted in workplace 
online communities. Tang et al.~\cite{tang} analyzed the difference in language usage of international Facebook users recently migrated to the United States to selectively self-disclose to their old (native) and new (English-speaking) social circles.
\subsection*{Lexical Compounding}
Lexical compounding constitutes an active area of research; there have been few studies on lexical compounding in English~\cite{bad,compound} and other languages like Italian, French, German, Spanish, Chinese etc.~\cite{chinese,gaeta,rocling,packard,french}. Pustejovsky~\cite{gl} provided one of the earliest explanation of the compounding phenomena within a compound based on the qualia modification relations in the semantic composition within a compound. A recent study by Lee et al.~\cite{rocling} discusses the formation of noun-noun compounds found in Chinese as well as few other languages like German, Spanish, Japanese and Italian. Word compounding is often termed as a form of lexical change which may be caused due to social pressure, ease of pronunciation. Hacken~\cite{pius} showed how translations can be used as heuristics to determine the concept of compounding. Noun-noun compounding is the most popular form and most studies are biased on restricting themselves to this form only. However, Bagasheva in~\
cite{alex} have studied the characteristics of compound verbs in English and 
Bulgarian language and claimed that verbs also compound to a significant extent. Bagasheva also showed that though the basic types of compound verbs are of the form verb-verb (blowdry, drinkdrive) and noun-verb (babysit, brainwash, proofread), other forms like noun-noun (handcuff, stonewall), adjective-noun (fastrack, badmouth), adjective-verb (whitewash, dryclean), preposition-noun/preposition-verb (overrun, underestimate) are also legitimate in English. We shall observe that similar kinds of POS (Parts of Speech) tag combinations are also present in case of hashtag compounds. In principle, we have attempted to merge the socio-linguistic features with information technological research. All of the above studies in linguistics considered anecdotal evidences by showcasing various examples and mostly attempted to study the formation of compounds. The main challenge of this type of research was the non-availability of temporal data of language change. We try to bridge the existing gap by considering large-scale 
social media data and study the compounding phenomena assuming hashtags as linguistic units. The previous studies on lexical compounds as discussed above have been mostly about understanding the formation of compounds whereas our analysis of the hashtag compounds is focused on the adoption of the compounds. There are differences in the mechanisms of compound formation in ordinary language and in social media. While lexical compounding is mostly spontaneous in nature, there are many purposes to which hashtag compounds are being put in social media. One aspect of it is market strategic hashtags like \#AmazonPrimeDay, \#CSCW2016 etc. Another reason of it is to fulfill communicative intents such as affective expression, political persuasion, or humor; for example hashtags like \#YesAllWomen, \#FeelTheBern, \#BlackLivesMatter, \#PresidentTrump etc. There is also spontaneous pressures of compounding like \#TheBestFeelingInARelationship, \#YouKnowItsRealWhen, \#RelationshipTips etc.

Our work has been inspired by several studies~\cite{cunha,zappa,paola,guardian} that focused on hashtags as linguistic units and attempted to identify the systematic similarities/differences with standard natural languages entities. Cunha et al.~\cite{cunha} studied hashtags as linguistic innovation and characterized the formation and usage of Twitter hashtags. Zappavigna~\cite{zappa} explored how hashtags enact three simultaneous communicative functions: marking experiential topics, enacting interpersonal relationships, and organizing text. Caleffi~\cite{paola} analyzed hashtagging as a productive process of word formation in English and Italian.

\subsection*{Lexical Blending}
Another closely related innovation phenomena is lexical blending where a word is formed from two or more words fused into one another. For example, brunch (breakfast + lunch), fantabulous (fantastic + fabulous), entertoyment (entertainment + toy) etc. This linguistic form of word reduction has been studied widely~\cite{brdar2008,carter2002,connolly2013,cook12,cook2010,gaskell1999,leturg,medler04,renner2012}. Gaskell and Marslen--Wilson~\shortcite{gaskell1999} have proposed a distributed model of speech perception for identification of blends, ambiguity etc. in spoken language. Cook and Stevenson~\shortcite{cook2010} have proposed a statistical model for identifying the lexical blend's source words from the observed linguistic properties of the blend. In a subsequent study, Cook~\shortcite{cook12} has proposed a regular expression based method for identifying lexical blends in social media. 

\subsection*{Hashtag popularity} Hashtags are a way for social media users to tag their posts with keywords, which in turn helps in meaningfully organizing the posts to make the contents on social networks easily searchable. Hashtags have various utilities. These are used in social campaigns, political campaigns, marketing and so on. They also provide great way to get people talking, and let them jump into discussions. For example, \#Polichat is a popular stream of conversation used by political and digital professionals. Therefore, it is important to know the popular and trending hashtags so that it is possible to filter out meaningful contents from the streams of data. There have been many studies on hashtag adoption (popularity)~\cite{kamath,kong,bigbird,sun,maon,what,Lilian2012srep,filippo1,yang}. Tsur and Rappoport in~\cite{what} performed content based prediction of hashtag popularity. Ma et al. in~\cite{maon} proposed a framework for predicting popularity of newly emergent hashtags. They showed that 
the contextual features based on the underlying social network of the users of the hashtag are more important than the content based features in predicting the popularity of a hashtag on a daily basis. Kamath and Caverlee~\cite{kamath} have modeled the geo-spatial propagation of online information spread to identify which hashtags will become popular in specific locations. Another notion of hashtag popularity is the ``burstiness'' of hashtag, the phenomena which involes sudden rise in hashtag usage and quick fall thereafter. Kong et al.~\cite{alex,kong} studied the 
burstiness of hashtag on a temporal scale. 

While retweets and followers support a hashtag's growth, they also paradoxically undermine its persistence. Various researchers have tried to systematically analyze the features that contribute to the growth and stabilization of the hashtags. Yang and Scott~\cite{Yang_predictingthe} examined the roles of ``relevance'' and ``exposure'' for hashtag adoption~\cite{Yang_predictingthe}. Yang et al.~\cite{yang} studied the duality of hashtags as topical identifiers and a symbol of community membership. Lin et al.~\cite{bigbird} studied the growth, survival, and context of novel
hashtags during the 2012 U.S. presidential debate. They proposed a framework to capture dynamics of hashtags based on their topicality, interactivity, diversity, and prominence.

\subsection*{Adoption and propagation of topical information in Twitter}
There have been several studies on the roles the users play in adoption and propagation of topical information in Twitter. Lerman and Ghosh~\cite{lerman} studied the user activities in Digg and Twitter. Lin et al.~\cite{lin} studied the evolution of a topic and revealed the diffusion path of the topic in the community. They proposed a probabilistic model based on textual documents, social influences of the users and topic evolution. Romero et al.~\cite{romero} observed that different topical category (sports, music, Idioms) of hashtags have different propagation pattern. Gomez et al.~\cite{gomez} studied the information diffusion among blogs and online news sources. Wu et al.~\cite{wu} analyzed the ``elite'' users and their roles in information spread. Starbird et al.~\cite{starbird} and Vieweg et al.~\cite{vieweg} addressed the characterization of events (natural hazards) by the Twitter users who posted tweets among them. Shamma et al.~\cite{shamma} proposed metrics for identification of nature of topics/
events (peaky or persistent) in Twitter data stream. De Choudhury et al.~\cite{munmun} characterized Twitter users into three primary categories: organizations, journalists/media bloggers, and ordinary individuals. Bhattacharya et al.~\cite{parantapa} characterized topical groups based on their network structures and tweeting behaviors. They distinguish between two types of users within a topical group; experts who are likely to be authoritative sources of information on specific topics, and seekers who are interested in gathering information on these topics.

Most of the above approaches in lexical compounding/blending propose theories/hypotheses with anecdotal evidences from various languages. However, we perform an in-depth large scale analysis of hashtag compounding phenomena and the adoption characteristics of the hashtag compounds in social media from publicly available data. We propose a prediction framework for early prediction of the popular hashtag compounds that is manifolds better than the baseline system based on human judgments. 
\section{Dataset description}
Twitter provides $1\%$ random sample of all the tweets via its sample API in real time. This API has been used to crawl tweets from $1^{st}$ July, 2011 to $31^{st}$ December, 2013. For analysis, we consider the users who have mentioned English as their language in their profile. We also performed a second level filtering of the tweets by a language detection software~\cite{langid} to remove any non-English tweets from the dataset. The data are then tokenized using the same tokenizer used by the CMU POS tagger~\cite{owu}. In total, the dataset consists of $\sim1$ billion tweets.

\section{Adoption of hashtag compounds} In this section, we discuss the phenomena of compounding and the adoption of hashtag compounds in social media. For detection of the hashtag compound, we consider 6 months data from $1^{st}$ January, 2012 to $30^{th}$ June, 2012. For any hashtag of length $\geq$ 6 in this data, of the form \#AB, we search for \#A and \#B in the data from $1^{st}$ July, 2011 till the time point when the first appearance of \#AB is found. Note that, both \#A and \#B themselves could be single words or compound words. For example, \#HighSchoolMemories (\#HighSchool + \#Memories), \#NeverShouldYouEver (\#NeverShould + \#YouEver) etc. We restrict our study to hashtag compounds that are formed by only two constituent hashtags (i.e., ignore further divisions of the constituent hashtags). To avoid ambiguity, we do not consider those compounds that can be formed due to the compounding of multiple pairs of constituent hashtags. For example, \#IStillHaventStarted (\#IStill + \#HaventStarted or \#
IStillHavent + \#Started).

Note that not all hashtag compounds become popular after the compounding. Out of $\sim2$ million candidate compounds, only $2\%$ are found to attain a frequency of usage more than both the constituent hashtags. In table~\ref{tab:posneg}, we show some examples of the compounds which are more frequently used than the constituent hashtags right from the point of the compound formation. In the same table, we also present examples of some compounds that do not gain frequency after the compound formation. Without loss of generality, we refer to the first type of hashtag compound as ``popular'' and the second type as ``unpopular''. We study in detail these two types of compound hashtags and their properties, thereby, identifying factors differentiating them. Understanding the precise reasons for certain compounds becoming popular could have far reaching impact both linguistically as well as in trending hashtag recommendation service where recommendation of a ``would-be-popular'' compound can be 
made even before such compounds are born.
\begin{table*}[ht]
\caption{Few examples of the popular and unpopular compound hashtags. The numbers in the parenthesis denote the frequency of the hashtag 10 months after the merging took place.}
\label{tab:posneg}
\begin{center}
\resizebox{16cm}{!}{
\begin{tabular}{|c|c||c|c|}\hline
Popular & Formation & Unpopular & Formation\\\hline
\#HighSchoolMemories (21700) & \#HighSchool (395) + \#Memories (4178) &      \#LoveOomf (1)& \#Love (14525) + \#Oomf (142299)\\
\#FreshmanAdvice (9144) & \#Freshman (102) + \#Advice (124) &      \#OomfPussy (5)& \#Oomf (142671) + \#Pussy (11010)\\
\#QuestionsIHateAnswering (4186) & \#QuestionsIHate (14) + \#Answering (1) &    \#ILovePorn (2) & \#ILove (428) + \#Porn (46715)\\
\#OperationLegalizeWeed (3978) & \#Operation (18) + \#LegalizeWeed (12)&        \#YOLOForJesus (1) & \#YOLO (47056) + \#ForJesus (4)\\
\#WikipediaBlackout (2638) & \#Wikipedia (202) + \#Blackout (524)&        \#HateCanada (3) & \#Hate (1622) + \#Canada (2399) \\
\#GameInsight (2633) & \#Game (689) + \#Insight (49) &    \#SweetBabyJesusThatsGood (1) & \#SweetBabyJesus (45) + \#ThatsGood (27)\\
\#CNNDebate (2615) & \#CNN (1637) + \#Debate (125)&        \#RegentStreet (1)& \#Regent (2) + \#Street (223)\\
\#GoldenGlobes (2581) & \#Golden (125) + \#Globes (61)&  \#ComingBackBlack (2)& \#ComingBack (12) + \#Black (1205)\\
\#GhettoSpellingBee (255) & \#Ghetto (134) + \#SpellingBee (8)&        \#LiquidationMonday (3)& \#Liquidation (51) + \#Monday (965)\\
\#LilWaynesGreatestHits (254) & \#LilWaynes (1) + \#GreatestHits (132)&        \#MavericksNation (4)& \#Mavericks (210) + \#Nation (136)\\ \hline
\end{tabular}
}
\label{tab:posneg}
\end{center}
\end{table*}
\subsection*{Popularity trend of the hashtag compounds} 
\begin{figure}[h]
\begin{center}
\includegraphics*[width=1\columnwidth,angle=0]{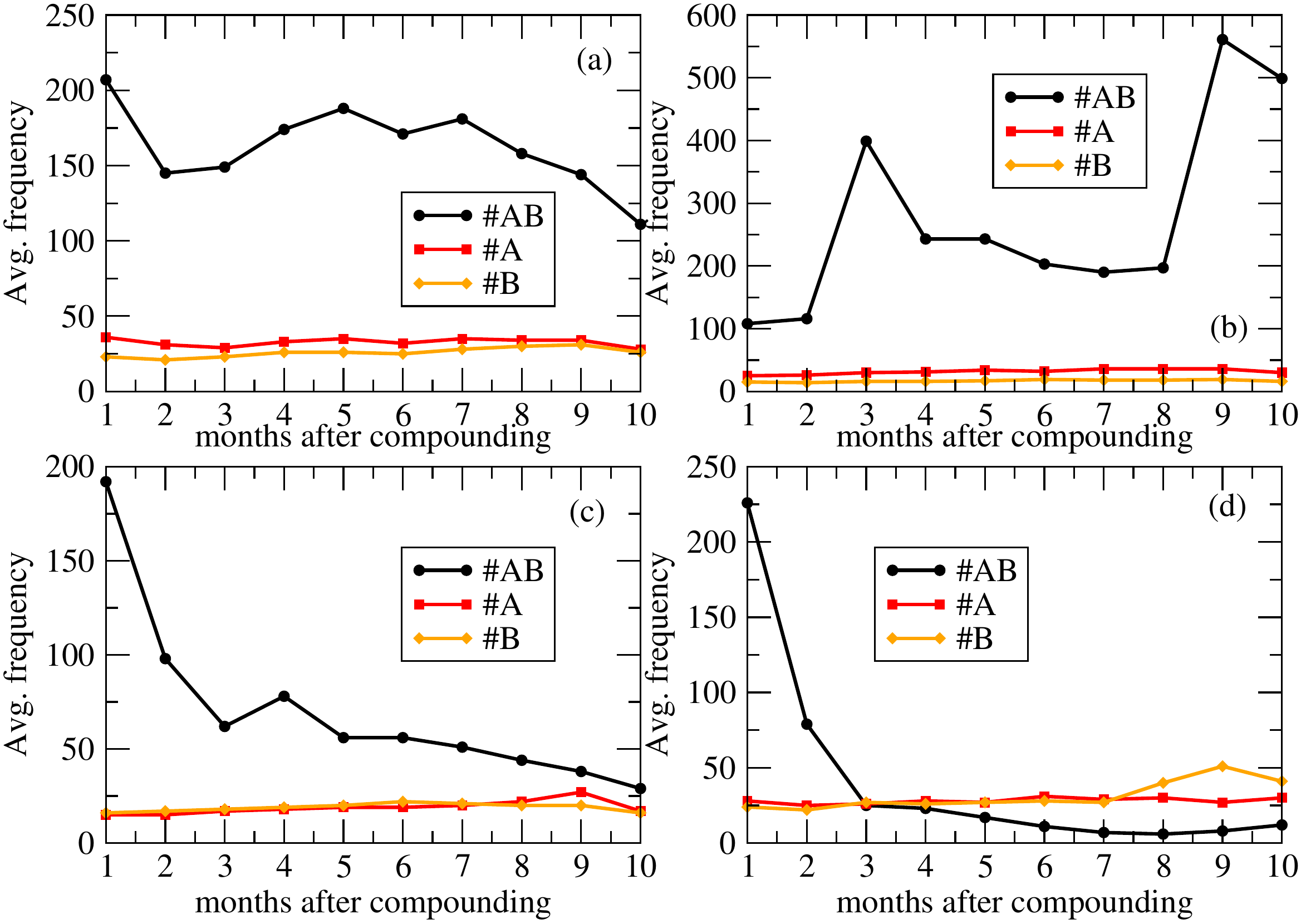}
\caption{\label{figtrend} The popularity trend of various categories of popular hashtags. The average frequency (no. of tweets) profile of the hashtags after the time of compounding for the hashtag triplets (\#A, \#B and \#AB) where the avg. frequency of the compounded hashtags \#AB are higher than both of the constituent hashtags \#A and \#B a) in all months b) in all but one month c) in all but two months d) in some months.}
\end{center}
\end{figure}

In this subsection, we shall discuss about the monthly popularity trend (frequency of the hashtag in tweets) of the popular hashtag compounds in the next 10 months after compounding. We categorize the popular hashtags into some finer classes - a) the frequency of the compound is always higher than that of its constituent ones b) the frequency of the compound is always higher compared to its constituent ones except for one month c) the frequency of the compound is always higher compared to its constituent ones except for two months d) the class containing the rest of the popular hashtag compounds. For the first three categories, the popularity trend suggests ``winner-takes-all''. We also observe that in many cases, though initially the frequency of the hashtag compounds remains higher than its constituent hashtag, however as time progresses the frequency falls below the constituent hashtags (see fig~\ref{figtrend}(d)). On an average, we see that for time 
larger than 2 months from the time point of compounding, this phenomenon takes place. This is the reason we select the first (early) prediction time point $T$ = 2 months (see section 7 for details).
\section{Linguistic aspects of hashtag compounding}
\label{ling}
This section is inspired by the observations made by researchers working on various aspects of lexical compounding. In addition to this, we also identify certain other interesting issues specific to hashtag compounding some of which might be as well generalized to mainstream lexical compounding research. In the rest of this section, we shall discuss various linguistic aspects of hashtag compound formation. We shall mostly focus on the compounding zone where the two constituent hashtags merge.
\subsection*{Part-of-speech combination} 
\begin{figure}[h]
\begin{center}
\includegraphics*[width=1\columnwidth,angle=0]{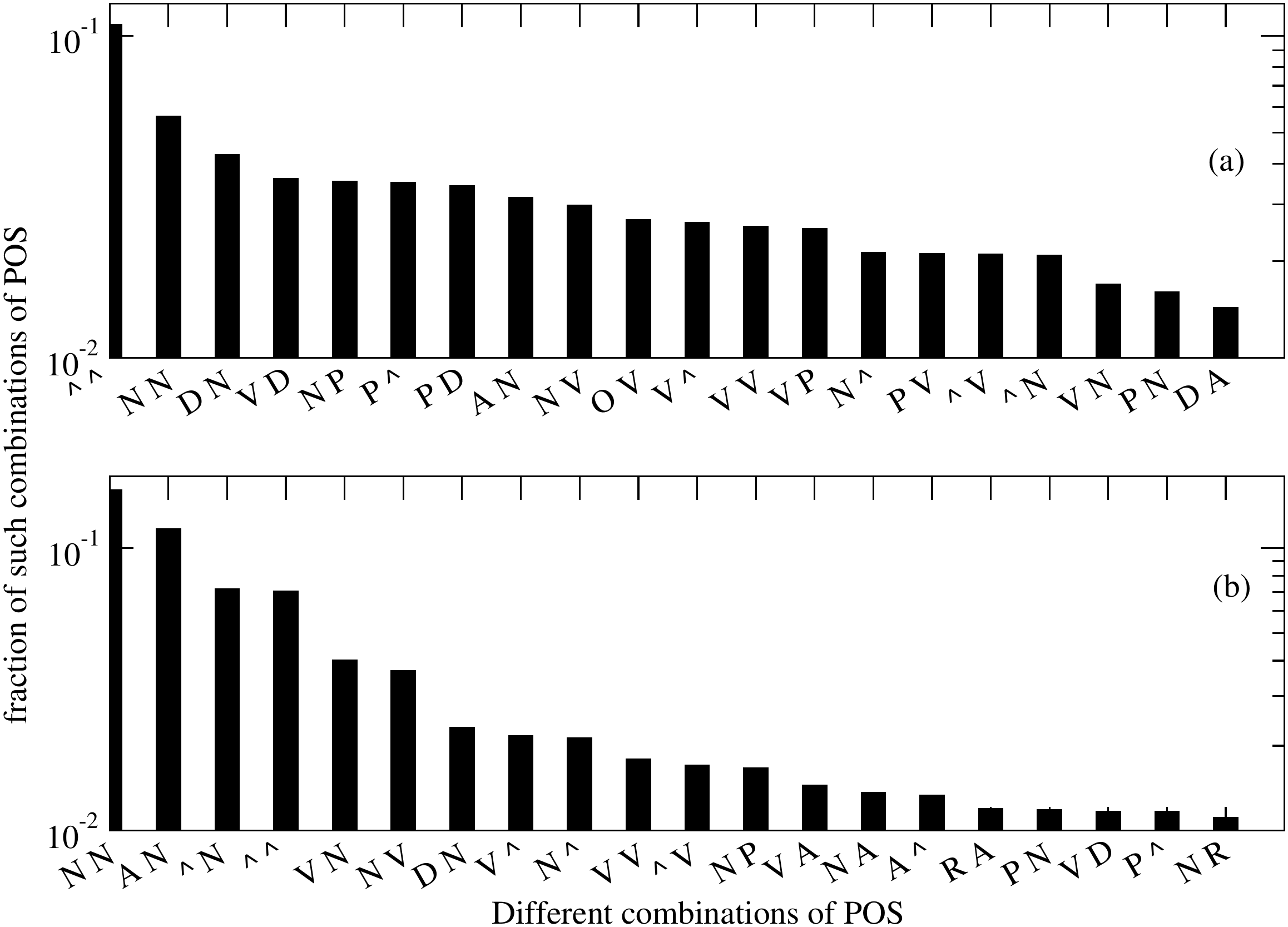}
\caption{\label{fig:pos} Distribution of top 20 most prominent combinations of part-of-speech tags of a) popular b) unpopular compound hashtags. The full form of the acronyms of the POS tags used here are as follows: \^{} - Proper Noun, $N$ - Common Noun, $O$ - Pronoun, $V$ - Verb, $A$ - Adjective, $R$ - Adverb, $D$ - Determiner, $P$ - Pre or post position}
\end{center}
\end{figure}

In lexical compounding, we find evidences of various types of compounds based on the POS~\cite{rocling} of the individual words that get compounded across various languages. For example, noun-noun, verb-noun, noun-verb, verb-verb etc are some common forms. We hypothesize that a similar phenomenon is instrumental in case of hashtags also. To validate this hypothesis we POS tag the individual hashtags using the CMU POS tagger~\cite{owu}, which is the state-of-the-art POS tagger available for Twitter. For a compound of the form \#AB which is made up of \#A and \#B, we find POS of \#A and POS of \#B to determine various combinations of POS-based compound formation. Note that the individual hashtags \#A and \#B can themselves be also compounds like \#ab and \#cd where $a$ and $b$ are words compounding to \#A and $c$ and $d$ are words compounding to \#B. In such scenarios, we consider the compounding zone and find the POS of the last part of \#A (i.e, $b$) and the POS of 
the first part of \#B (i.e., $c$). In figure~\ref{fig:pos}, we show the distribution of various POS-combinations of the hashtag compounds. We observe that there is a clear distinction present between the distribution of POS combinations for popular and unpopular compounds. Most prominent POS combinations in case of popular compounds are proper noun-proper noun, followed by common noun-common noun, determiner-common noun and verb-determiner; however, the most prominent POS combinations for the unpopular hashtag compounds are common noun-common noun, Adjective-common noun, determiner-common noun, proper noun-proper noun etc. Among both popular and unpopular compounds, the common noun-common noun pair seems to be very prevalent.

\subsection*{Named entity combination}
\begin{figure}[ht]
\begin{center}
\includegraphics*[width=1\columnwidth,angle=0]{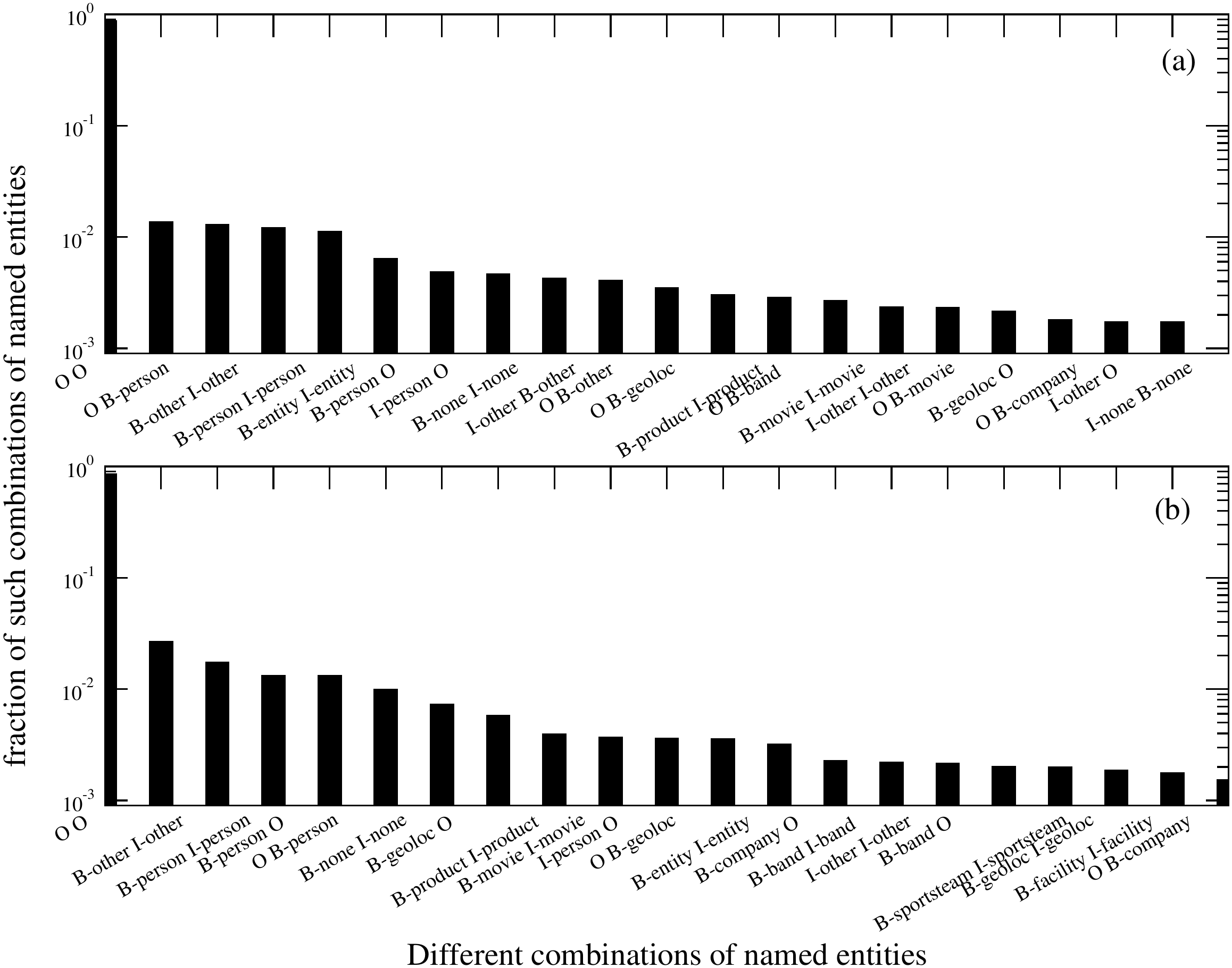}
\caption{\label{fig:namedentity} Distribution of top 20 most prominent combinations of named entities of a) popular b) unpopular compound hashtags. For detailed description of the named entity types, refer to \url{https://github.com/aritter/twitter_nlp}}
\end{center}
\end{figure}

Apart from the POS tags, we also perform named entity recognition of the constituent hashtags forming the compound to understand which types of entities merge. We use a named entity recognition tool~\cite{ritter} for identifying named entities of the words in the hashtags forming the compounds. For a compound hashtag (\#AB = \#A + \#B where \#A = \#ab and \#B = \#cd), we find the named entity of last word of \#A and the named entity of first word of \#B to determine various combinations of named entity-based compounding. For other cases where \#A, \#B are single words we perform the recognition directly on these words. Figure~\ref{fig:namedentity} shows the distribution of the top 20 most prominent named entity combinations for popular (fig~\ref{fig:namedentity}(a)) as well as unpopular hashtag compounds (fig~\ref{fig:namedentity}(b)). Though in 85\% cases we find that the constituent words are non-entities, from the remaining 15\% cases, we find various named entity combinations and the distribution of 
these 
combinations are indeed very different for the popular and the unpopular classes. In general, the most prevalent named entity combinations where both the constituent hashtags denote entities are (B-person I-person) followed by (B-product I-product) and (B-movie I-movie). However, the fraction of each such pair for the popular compounds is very different for the unpopular ones.

\subsection*{Out-of-vocabulary / In-vocabulary combination} 
With the advent of new words/slangs in social media, there is an increasing trend of usage of out-of-vocabulary (OOV) words~\cite{eisenstein}. Motivated by above observation, we study if OOV words have a role in compound formation. We use GNU Aspell dictionary to determine whether a given word is an OOV or INV (In-Vocabulary). As stated earlier, for each compound hashtag of the form \#AB formed by \#A and \#B, we find the nature of the ending word of \#A and the beginning word of \#B. In table~\ref{tab:oov}, we report the distribution of various combinations for both the popular and unpopular compounds. The most prevalent combination in both cases are the merge of two INV words (though varying in the percentages, $\sim44\%$ in case of popular compounds compared to $\sim67\%$ in case of unpopular compounds). There is also a marked distinction in the rank order in which the combinations are used apart from usage variability. The rank order in case of popular compounds is OOV-OOV, INV-OOV, OOV-INV whereas for 
the unpopular ones it is OOV-INV, INV-OOV, OOV-OOV.
\begin{table}[ht]
\begin{center}
\small
\caption{Distribution of various combinations of OOV and INV words of popular and unpopular compound hashtags.}
\label{tab:oov}
\begin{tabular}{|p{1.6cm}|p{1.2cm}|p{1.6cm}|p{1cm}|}\hline
\multicolumn{2}{|c|}{Popular}& \multicolumn{2}{c|}{Unpopular} \\ \hline
Combinations & \% & Combinations & \%\\\hline
INV-INV & 43.9& INV-INV&66.9\\ \hline
OOV-OOV & 20.7 & OOV-INV & 14.0 \\ \hline
INV-OOV &19.8&INV-OOV&13.6 \\ \hline
OOV-OOV & 15.6 &OOV-OOV &5.5 \\ \hline
\end{tabular}
\end{center}
\end{table}

\section{Baseline system} The main purpose of the human prediction is to find out whether humans can identify the popular compound just from the structural information of the hashtags. If humans can easily identify popular compounds then the whole problem of predicting popular hashtag compounds is not interesting. The purpose of this work is to design an automated framework which will assist humans adopting hashtag compounds that are going to be popular in near future and we shall compare how good this system performs by considering human judgment as baseline. 
To understand whether humans can predict if a compound is going to be popular in future, we conduct an online survey\footnote{\url{http://bit.ly/1ARJlRp}} among $72$ agreed participants (students, researchers, professors, technical persons) with ages ranging between 18-34 years. We choose 600 hashtag compounds randomly from the set of 2000 compounds used for classification (see section 8). Each participant is given a set of 25 questions. In each question, the participants are given the hashtag compound as well as the constituent hashtags and are asked whether the compound hashtag would become more popular in future than both the constituent hashtags. If they are not sure of the answer, they have the option to indicate that they do not know. Each question is asked to exactly 3 participants. A detailed analysis of the survey results is outlined below. We receive $\sim15\%$ responses where the participants indicated that they are unsure. Out of remaining 85\% responses, 53.3\% responses 
are found to be correct answers. We adopt majority voting technique to evaluate each question. $\sim10.5\%$ of the questions remain undecided due to all the possible answers getting equal number of votes. Out of the remaining questions, we obtain an accuracy of 54.5\% and an overall accuracy of $48.7\%$. We also perform averaging of responses. For each hashtag compound, we find the fraction of responses in agreement with the real data. Then, we take average for all the hashtag compounds. This yields an overall accuracy of $45.33\%$. To find out the inter-evaluator agreement, we compute Fleiss' Kappa~\cite{fleiss} which is found to be 0.15. In order to identify how the individual user judgments are, we compute response accuracies for each user separately. The median user response accuracy comes out to be 44\% and the standard deviation of the user response accuracies is 0.12. The maximum and minimum user response accuracies are 68\% and 12\% respectively.
\begin{table}[ht]
\begin{small}
\caption{Baseline accuracies.}
\label{tab:manual}
\begin{center}
\begin{tabular}{|c|c|} \hline
Method & Overall Accuracy \\ \hline
Majority Voting & 48.7\% \\ \hline
Averaging & 45.33\% \\ \hline
\end{tabular}
\end{center}
\end{small}
\end{table}

From the above observations (table~\ref{tab:manual}) and the discussions, we can conclude that human judgment is on an average poor. This motivates us to develop an automated prediction framework which as we shall see is highly accurate. We consider the human judgment accuracies as a baseline for our prediction model presented in the next section.
\section{Prediction model}
\begin{figure*}[!t]
\begin{center}
\includegraphics*[scale = 0.4, angle=0]{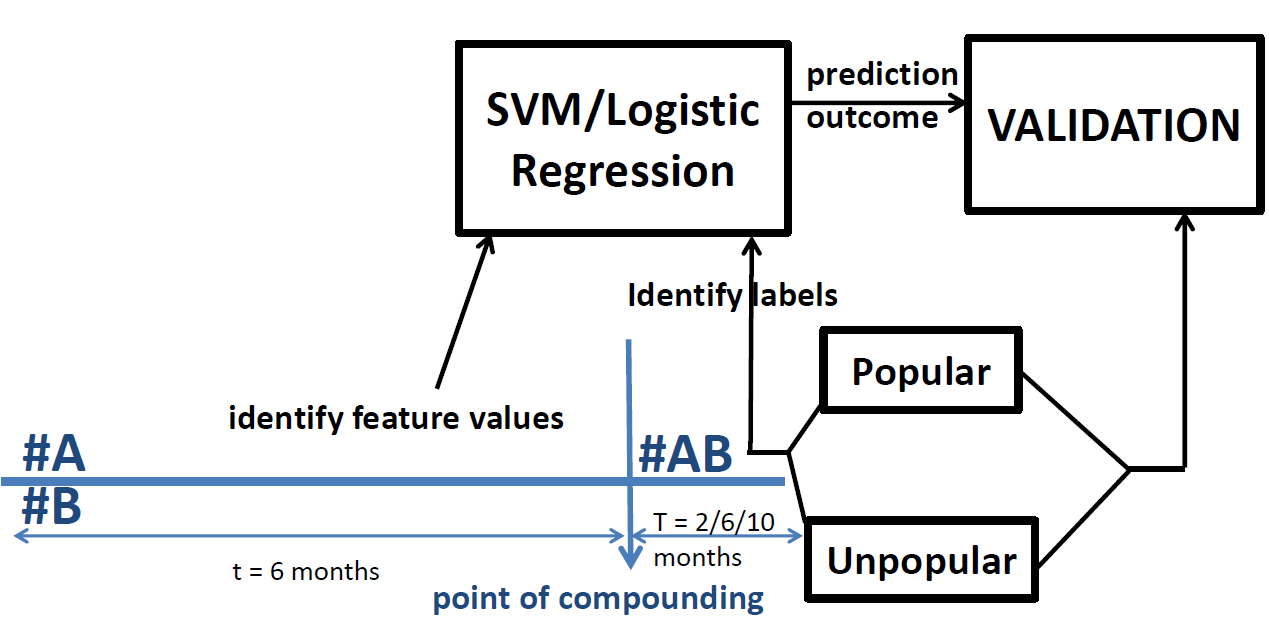}
\caption{\label{figflow}A schematic of our proposed framework. Here the popular category refers to those cases where frequency of usage of \#AB $>$ frequency of usage of \#A AND frequency of usage of \#B. The rest of cases are categorized as unpopular.}
\end{center}
\end{figure*}

In this section, we propose a model for early prediction of the ``would-be trending'' hashtag compounds. For the prediction task, we observe the constituent hashtags (\#A, \#B) for $t$ = 6 months before they get compounded together to form \#AB and predict whether \#AB will be more popular (in terms of frequency in tweets) than both \#A and \#B or not after $T$ months from the time point of the compounding (see fig~\ref{figflow}). In our setting, we consider $T$ = 2, 6 and 10 months. 

For the task of prediction, we learn three major types of features : \\
\textbf{Hashtag content features} - the features that are related to the content of the hashtag only \\
\textbf{Tweet content features} - the features that are related to the tweets in which the hashtags appear \\
\textbf{User features} - these include various properties of the users who tweet the hashtags, such as their social influence etc.
\subsection*{Hashtag content features:}
For each hashtag we extract various attributes related to its content. These are mostly the attributes related to characters, words and the nature of the words that are used in the hashtag.

\subsubsection*{Character length of the compound hashtag} Due to the 140 character-limit on the tweets, character usage is vital and is hence a constraint on the size of the hashtags too. People tend to express their feeling using smaller number of characters but there is a trade-off; smaller sized hashtags do not always serve their purpose. Therefore, the number of characters in the hashtag compound is a feature of the model.

\subsubsection*{Number of words in the compound hashtag} The number of the words in a compound hashtag is also important because more words may mean more expressibility. The compound hashtag might be more expressible than the constituent hashtags.

\subsubsection*{Presence of n-grams in English texts} We segment the words in the compound hashtag and search for 2, 3, 4, 5 grams of the constituent words in the corpus of 1 million contemporary American English words\footnote{http://www.ngrams.info/samples\_coca1.asp}. We use presence of any of these n-grams as a feature for the classifier. 

\subsubsection*{Part-of-speech diversity of the words in the compound} We use standard CMU POS tagger~\cite{owu} as stated earlier for identifying the POS tags after segmenting the compound hashtag into its constituent words. We define the POS diversity (POSDiv) as follows:
\begin{equation*}
 POSDiv (h_i) = -\sum_{j \in P}p_j\times\log(p_j)
\end{equation*}
where $h_i$ is the $i^{th}$ hashtag and $p_j$ is the probability that a word is labeled by the $j^{th}$ POS from the set $P$ of all possible POS tags. We use this diversity metric as a feature for our classifier.

\subsubsection*{Part-of-speech combination} As observed earlier in section 5, there is clear distinction in the distribution of POS tag combination for the popular and the unpopular compounds in the compounding zone. Motivated by this observation, we consider the POS tag combinations as features to the classifier. We consider 20 such most prevalent POS tag combination, each of them acting as a binary feature.

\subsubsection*{Named entity combination} Similar to the above feature, we also consider the named entity combination as an important feature for the classifier. In section 5, we observe that the most prominent named entity combinations are different for the popular and the unpopular hashtag compounds. Therefore, we consider 20 most prominent named entity combinations, each one of them as a binary feature for the classifier.

\subsubsection*{OOV/INV combination} We also observe in section 5 that there are significant differences in the distribution of the various INV/OOV combination for the words at the point of merge. To utilize this striking difference, we consider all four combinations (OOV-OOV, INV-OOV, OOV-INV, INV-INV) as 4 individual binary features for the classifier.

\subsection*{Tweet content features:}
The content of tweets that use a hashtag is also a significant determinant of the popularity of a hashtag compound. In this subsection, we shall be describing a series of tweet content features.

\subsubsection*{Word overlap} We compute the overlap coefficient\footnote{overlap coefficient between two sets $A$ and $B$ is given by $overlap (A, B) = \frac{|A \cap B |}{min(|A|,|B|)}$} between the set of words appearing in tweets with \#A and \#B. This overlap coefficient act as a feature for our classifier.

\subsubsection*{n-gram overlap} For the compounding hashtags \#A and \#B, we consider the words appearing in tweets with those hashtags separately and search for 2, 3, 4, 5 grams in the corpus of 1 million contemporary American English words\footnote{http://www.ngrams.info/samples\_coca1.asp}. We then find out the overlap coefficient between the set of valid n-grams for \#A and \#B.

\subsubsection*{Average frequency of the overlapping set of n-grams} In a similar way as above, we find out the average frequency of the n-grams in the contemporary American English corpus for the overlapping set of n-grams found from the set of tweets for \#A and \#B.

\subsubsection*{Collocation frequency of the compounding pair} To understand whether collocation of the compounding hashtags in tweets has effect on the compound formation thereafter, we consider the collocation frequency of the compounding hashtags as feature for the classification task.

\subsubsection*{Clarity of the compounding hashtags} Hashtag clarity, a metric that has been defined in~\cite{sun} quantifies topical cohesiveness of all the tweets in which the hashtag appears. Clarity of hashtag $i$ ($HashClarity(i)$) is computed as the Kullback-Leibler (KL) divergence between the unigram language model inferred from the document $D_i$ containing all the tweets for the $i^{th}$ hashtag and the background language model from the entire tweet collection $T$. If a hashtag refers to a specific topic, then the high probabilities of a few topic-relevant words distinguish its tweets from the background.
\begin{equation*}
 HashClarity(i) = -\sum_{w \in D_i}p(w|D_i)\times\log\frac{p(w|D_i)}{p(w|T)}
\end{equation*}
We compute hashtag clarity for both the compounding hashtags \#A and \#B and use them individually as features.

\subsubsection*{Word diversity of the compounding hashtags} This feature tells us how much diverse are the words related to a hashtag. If $D_i$ is the document containing all the tweets in which hashtag $i$ appears and $p(w|D_i)$ is the probability of a word belonging to the document $D_i$ then word diversity of hashtag $i$ is defined as follows
\begin{equation*}
 WordDiv(i) = -\sum_{w \in D_i}p(w|D_i)\times\log p(w|D_i)
\end{equation*}
We compute word diversity of both the compounding hashtags \#A and \#B and use each of them as a feature.

\subsubsection*{Avg. topic overlap among the compounding hashtags} Topical overlap is an important aspect for a hashtag compounding phenomena. More the constituent hashtags are aligned to the similar topics, more is the chance that the hashtag compound becomes popular. For topic discovery from the tweet corpus, we adopt Latent Dirichlet Allocation (LDA)~\cite{lda} model, a renowned generative probabilistic model for discovery of latent topics. For a hashtag $i$, we consider all the tweets in which the hashtag appears as a document for the LDA model. Now, considering all the hashtags we experiment with, we have a collection of documents on which we run LDA to obtain the word distribution across all the topics for each document. Next, for each of the topic, we find top 100 words according to the belongingness probability of the words in the topic for both \#A and \#B. We then compute the overlap between these two sets. For each topic, we compute topic overlap (in terms of the number of common words belonging 
to that topic) between \#A and \#B and consider the average of them as a feature for the classification model.

\subsection*{User features:} Users play an important role in hashtag adoption. People use/adopt hashtags according to their personal interest, their social influence etc. In this subsection, we shall discuss a set of user features which could be important for discriminating a popular compound from an unpopular one.

\subsubsection*{Unique and common users} We hypothesize that the extent of adoption of the constituent hashtags could be an important indicator of the overall popularity of the compound formed. For this reason, we measure the number of unique users tweeting using either of the constituent \#A or \#B. These two counts act as classification features. In addition, we also identify the number of common users who tweet both the constituent hashtags \#A and \#B either in the same tweet or in different tweets. This is another feature for the classification model.

\subsubsection*{Mention behavior of the users} People tend to mention people in tweets whom they like to engage in conversations. Thus, mention behavior in tweets for constituent hashtags might affect adoption of the hashtag compound. Therefore, we find the number of unique users mentioned in tweets containing the constituent hashtag \#A. The same is found for \#B. These two act as features for the classifier. We further find the number of common users being mentioned either in the same or different tweets containing both the constituent hashtags \#A and \#B. This is another feature for the classification model.

\subsubsection*{Retweet behavior of the users} Retweeting is an inherent indicator of increasing popularity of a hashtag. More retweets usually mean more popularity. Similar to the case of mentions, we find the number of unique retweets for the set of  tweets containing the constituent hashtag \#A. The same goes for \#B. These two act as features for the classification model. We thereafter find number of common retweets using both the compounding hashtags \#A and \#B. This is also a classification feature.

\section{Performance evaluation}
In this section, we analyze the performance of our prediction model. For prediction task, we use 2000 hashtag compounds (\#AB) whose constituent hashtags \#A and \#B have each occurred in at least 50 tweets six months before the time point at which the compounding took place. We use Support Vector Machine (SVM) and logistic regression classifiers available in Weka Toolkit~\cite{weka} for classifying the data into popular and unpopular hashtag compounds. We perform 10-fold cross-validation as well as training and testing on separate dataset by splitting the data into 9:1 (see table~\ref{tab:classification} for details). We achieve 77.07\% accuracy with high precision and recall rates while predicting after $T$ = 2 months. As one increases this time period, the accuracy of prediction increases, although not very significantly. For long-term predictions after $T$ = 6 months and 10 months, the accuracy obtained are 77.5\% and 79.13\% respectively. Both the classifiers yield very similar classification 
performance; 
however the logistic regression model gives better area under the ROC curve compared to SVM. We also observe that the number of topics ($K$) of LDA do not have a significant effect on the classification results. For $K = 30$, we achieve the best accuracy and the area under the ROC curve. We observe that our prediction accuracy improves by $\sim$58\% over the baseline accuracy produced by human judgment.
\begin{table}[!t]
\small
\centering
\caption{Performance of various classifiers at different time of prediction ($T$ = 2, 6, 10 months) for different topic selection for LDA feature with number of topics ($K$ = 10, 20, 30, 40, 50). The classification results are shown for 10-fold cross validation as well as with separate training and testing set in 9:1 ratio.}
\label{tab:classification}
 \begin{tabular}{ |p{0.7cm}|p{1.3cm}|p{0.2cm}|p{0.6cm}|p{0.6cm}|p{0.6cm}|p{0.6cm}|p{0.6cm}| }
\hline
Time period & Classifier & $K$ & Accur-acy &Preci-sion & Recall & F-Score & ROC Area \\ \hline
 \multirow{12}{0.7cm}{$T$ = 2 months} & \multirow{5}{1.3cm}{SVM\tiny(10-fold cross validation)} & 10 & 76.18 & 0.762 & 0.762 & 0.762 &  0.762\\ \cline{3-8}
 & & 20 & 76.42 & 0.764 & 0.764 & 0.764 &  0.764\\ \cline{3-8}
 & &\textbf{30} & \textbf{77.07} & \textbf{0.771} & \textbf{0.771} & \textbf{0.771} &  \textbf{0.771}\\ \cline{3-8}
 & & 40 & 76.37 & 0.764 &  0.764 &  0.764 &   0.764\\ \cline{3-8}
 & & 50 & 76.72 & 0.767 &  0.767 &  0.767 &   0.767\\ \cline{2-8}
 & \multirow{5}{1.3cm}{Logistic Regression\tiny(10-fold cross validation)} & 10 & 76.13 & 0.761 & 0.761 & 0.761 &  0.836\\ \cline{3-8}
 & & 20 & 76.43 & 0.764 & 0.764 & 0.764 &  0.839\\ \cline{3-8}
 & & \textbf{30} & \textbf{76.48} & \textbf{0.765} & \textbf{0.765} & \textbf{0.765} &  \textbf{0.841}\\ \cline{3-8}
 & & 40 & 76.27 & 0.763 &  0.763 &  0.763 &   0.838\\ \cline{3-8}
 & & 50 & 76.42 & 0.764 &  0.764 &  0.764 &   0.837 \\ \cline{2-8}
 & SVM\tiny(seperate train and test set) & 30 &77.7 &0.777 &0.77 & 0.772&0.771 \\ \cline{2-8}
 & Logistic Regression\tiny(seperate train and test set) & 30 &77.5 & 0.781&0.775 &0.776 &0.834 \\ \hline \hline

 \multirow{12}{0.7cm}{$T$ = 6 months} & \multirow{5}{1.3cm}{SVM\tiny(10-fold cross validation)} & 10 & 76.85 & 0.769 & 0.768 & 0.768 &  0.768\\ \cline{3-8}
 & & 20 & 77.07 & 0.771 & 0.771 & 0.771 &  0.771\\ \cline{3-8}
 & &\textbf{30} & \textbf{77.52} & \textbf{0.775} & \textbf{0.775} & \textbf{0.775} &  \textbf{0.775}\\ \cline{3-8}
 & & 40 & 77.18 & 0.772 &  0.772 &  0.772 &   0.772\\ \cline{3-8}
 & & 50 & 76.4 & 0.764 &  0.764 &  0.764 &   0.764\\ \cline{2-8}
 & \multirow{5}{1.3cm}{Logistic Regression\tiny(10-fold cross validation)} & 10 & 75.84 & 0.758 & 0.758 & 0.758 &  0.817\\ \cline{3-8}
 & & 20 & 75.95 & 0.76 & 0.76 & 0.759 &  0.821\\ \cline{3-8}
 & & \textbf{30} & \textbf{76.62} & \textbf{0.766} & \textbf{0.766} & \textbf{0.766} &  \textbf{0.823}\\ \cline{3-8}
 & & 40 & 76.17 & 0.762 &  0.762 &  0.762 &   0.82\\ \cline{3-8}
 & & 50 & 75.84 & 0.758 &  0.758 &  0.758 &   0.819 \\ \cline{2-8}
 & SVM\tiny(seperate train and test set) & 30 &80 &0.832 &0.8 &0.802 &0.819 \\ \cline{2-8}
 & Logistic Regression\tiny(seperate train and test set) & 30 & 78.89&0.817 &0.789 &0.791 & 0.888 \\ \hline \hline

\multirow{12}{0.7cm}{$T$ = 10 months} & \multirow{5}{1.3cm}{SVM\tiny(10-fold cross validation)} & 10 & 76.7 & 0.768 & 0.767 & 0.767 &  0.767\\ \cline{3-8}
 & & 20 & 78.48 & 0.786 & 0.785 & 0.785 &  0.785\\ \cline{3-8}
 & &\textbf{30} & \textbf{79.13} & \textbf{0.792} & \textbf{0.791} & \textbf{0.791} &  \textbf{0.791}\\ \cline{3-8}
 & & 40 & 77.83 & 0.78 &  0.778 &  0.778 &   0.778\\ \cline{3-8}
 & & 50 & 77.02 & 0.772 &  0.77 &  0.77 &   0.77\\ \cline{2-8}
 & \multirow{5}{1.3cm}{Logistic Regression\tiny(10-fold cross validation)} & 10 & 77.7 & 0.777 & 0.777 & 0.777 &  0.824\\ \cline{3-8}
 & & 20 & 78.31 & 0.784 & 0.783 & 0.783 &  0.827\\ \cline{3-8}
 & & \textbf{30} & \textbf{78.65} & \textbf{0.787} & \textbf{0.787} & \textbf{0.786} &  \textbf{0.833}\\ \cline{3-8}
 & & 40 & 77.99 & 0.781 &  0.78 &  0.78 &   0.828\\ \cline{3-8}
 & & 50 & 78.6 & 0.786 &  0.786 &  0.786 &   0.827 \\ \cline{2-8}
 & SVM\tiny(seperate train and test set) & 30 & 79.03& 0.79&0.825 &0.79 &0.791 \\ \cline{2-8}
 & Logistic Regression\tiny(seperate train and test set) & 30 & 77.42& 0.816& 0.774&0.774 &0.892 \\ \hline
  \end{tabular}
\end{table}
\subsection*{Ablation experiments for feature importance} To understand the importance of the features, we perform ablation experiments by removal of various feature types. In table~\ref{feature_combo}, we present the contribution of different combinations of feature types, demonstrating how each of these feature types affect the classification and whether any feature type is masked by a stronger signal produced by other feature types. We observe that tweet content features are the most discriminative ones whereas hashtag content features are the least. However, the combination of tweet content features with user features and tweet content features with hashtag content features yield very similar accuracy values.
\begin{table}[h]
\small
\centering
\caption{Performance of various combinations of feature categories for $K=30$ and time period of observation $t$ = 2 months.}
  \label{feature_combo}
 \begin{tabular}{|p{5cm}|p{1.5cm}|}
\hline
Feature model & Accuracy \\ \hline
\textbf{All} & \textbf{77.07\%}\\ \hline
tweet content + user & 75.9\%\\ \hline
tweet content + hashtag content&75.12\% \\ \hline
hashtag content + user&72.4\% \\ \hline
tweet content &74.1\% \\ \hline
user& 68.18 \%  \\ \hline
hashtag content &65.04\% \\ \hline
  \end{tabular}
\end{table}
\subsection*{Discriminative features}
In this subsection, we discuss about the discriminative power of the individual features. In order to determine the discriminative power of each feature, we compute the chi-square ($\chi^2$) value and the information gain. Table~\ref{tab:featurerank} shows the order of all features based on the $\chi^2$ value, where larger the value, higher is the discriminative power. The ranks of the features are very similar when ranked by information gain (Kullback-Leibler divergence). Among tweet content features, the most discriminative features are the overlap features like n-gram overlap, word overlap. In addition, we observe that the hashtag features like the INV/OOV combination, POS combinations, are also highly discriminative.
\begin{table}[!htb]
\caption{Top 30 predictive features and their discriminative power for $K = 30$.}
\label{tab:featurerank}
\centering
\resizebox{7.5cm}{!}{
 \begin{tabular}{ |p{0.7cm}|p{1cm}|c| }
\hline
Rank & $\chi^2$ Value & Feature  \\ \hline
1&476.49& n-gram overlap \\ 
2&460.34& Avg. frequency of common n-grams \\
3&420.99& Word overlap \\
4&314.13& no. of unique retweets with \#A \\
5&285.79& no. of unique users tweeting \#A \\
6&281.71& no. of unique retweets with \#B \\
7&275.42& no. of unique users tweeting \#B \\
8&273.04& Word diversity of \#A \\
9&252.92& Hashtag clarity of \#A \\
10&222.32& Word diversity of \#B \\
11&213.82& no. of unique users mentioned in \#A \\
12&208.41& no. of unique users mentioned in \#B \\
13&204.14& Hashtag clarity of \#B \\
14&152.38& INV-INV \\
15&136.74& OOV-OOV \\
16&111.69& POS diversity \\
17&105.52& no. of common users tweeting using \#A and \#B \\
18&101.47& Avg. topic overlap \\
19&86.39& total no of words in \#AB \\
20&84.20& no. of common retweets for \#A and \#B \\
21&70.08& AN \\
22&65.22& OO \\
23&46.13& \^{}\^{} \\
24&39.99& no. of characters in \#AB \\
25&25.42& no. of common users mentioned for \#A and \#B \\
26&23.2& NN \\
27&16.13& B-personI-person \\
28&13.55& P\^{} \\
29&9.51& OOV-INV \\
30&9.04& AA \\ \hline
\end{tabular}
}
\end{table}

\section{Correspondence analysis} In this section, we shall compare the outcomes of the automatic prediction framework with the human judgment results. We consider this correspondence analysis to be a methodological novelty of our work and argue that such a study can form a crucial part of any future research of similar type.

For this purpose, we select from among the set of 2000 hashtags those 600 cases that have been used for the human judgment experiments. This time we train the model on the 1400 (i.e, 2000 - 600) cases and test on the 600 cases ($T$ = 6 months). We then compare the predicted labels from the automatic prediction framework and the human judgment labels decided via majority voting. In table~\ref{concord}, we present the results of this correspondence analysis. The number of discordant cases are higher than the number of concordant cases. We further observe that there are 211 cases where both human and automatic prediction framework correctly identify the labels; 246 cases where the automatic prediction framework is only able to identify the correct labels and humans fail to do so; finally there are non-significant number of cases (only 80) where humans could correctly identify the labels while the automatic prediction framework failed to do so.  
\begin{table}[h]
\small
\centering
\caption{Correspondence analysis}
  \label{concord}
 \begin{tabular}{|p{5cm}|p{1.5cm}|}
\hline
Concordance & 257 \\ \hline
Discordance & 343\\ \hline
Correctly judged by both human and automatic framework & 211\\ \hline
Wrongly judged by both human and automatic framework&46 \\ \hline
Correctly judged by only human&80 \\ \hline
Correctly judged by only automatic framework&246\\ \hline
  \end{tabular}
\end{table}
In table~\ref{concordreason}, we attempt to present reasons behind the above observations. We find that the human evaluators can correctly label those cases where the hashtag compound have the highest frequency for the popular class and lowest for the unpopular class (i.e., the relatively easier cases); on the other hand, the automatic prediction framework can identify the popular hashtag compounds whose frequency values are not very different from the constituent hashtags (i.e., a relatively harder case).
\begin{table}[t]
\small
\centering
\caption{ Cause analysis for the correspondence. The cell values represent the average frequency of the corresponding hashtag types. $HE$ : Human evaluator, $AF$ : Automatic framework}
  \label{concordreason}
  \resizebox{8.5cm}{!}{
 \begin{tabular}{ p{3cm}|p{1cm}|p{1cm}|p{1cm}|p{1cm}|p{1cm}|p{1cm}|}
\cline{2-7}
 & \multicolumn{3}{|c|}{Popular} & \multicolumn{3}{c|}{Unpopular}\\ \cline{2-7}
 & \#AB & \#A &\#B & \#AB & \#A &\#B \\ \hline
\multicolumn{1}{ |p{3cm}| }{Correctly judged by both $HE$ and $AF$} & 1324.25 & 130.63 & 117.62 & 2.4 & 1369.34 & 1297.47\\ \hline
\multicolumn{1}{ |p{3cm}| }{Wrongly judged by both $HE$ and $AF$}&1610.33 & 433.25 &136.58 & 5.77 & 180.7 &250.23 \\ \hline
\multicolumn{1}{ |p{3cm}| }{Correctly judged by only $HE$}&1644.4 &460.24 &332.5 & 1.48 &576.05 & 949.33 \\ \hline
\multicolumn{1}{ |p{3cm}| }{Correctly judged by only $AF$}&259.3&46.3&73.3&12.24&1130.91&849.34\\ \hline
  \end{tabular}}
\end{table}

\section{Discussions}
In this section, we shall discuss some of the implications of the findings from our study. The novelty of our research is tied to the method of merging socio-linguistic features with information technological research. We used hashtags as linguistic unit. Similar to lexical compounding, hashtag compounding also exhibit prominence of noun-noun compounding. Unlike lexical compounds, there are several other forms of compounding prevalent in popular hashtag compounds like determiner-noun compound, verb-determiner compound. We perform named entity recognition to identify which kind of named entities merge. We observe that combinations where both hashtags are entities, are predominantly found to be person-person combination and product-product combination whereas other kind of combinations are also prevalent. To the best of our knowledge, entity combination have not been studied in lexical compounding and hence is a novel contribution. Apart from the linguistic aspects of compounding, there are 
sociological factors which mostly drive the adoption process of compounds in communities. Due to unavailability of large scale data, there have been no prior work on this. In our work, we attempt to study the sociological aspects of hashtag compounding on large-scale Twitter dataset. We observe that mention and retweeting behavior of individuals are important factors for popular hashtag compounding. These features appear to be highly discriminative in predicting popular hashtag compounds. We also compare our automatic prediction framework 
with human judgment in order to justify the hardness of the prediction task. The framework can guide Twitter users in selecting the right compounds leading to a higher gain in popularity. We also perform a correspondence analysis of human judgment and machine prediction to find out whether there is any inherent pattern embedded in it. We find that human evaluators can guess the relatively easier cases where there are larger frequency gaps in a compound and its constituent parts whereas the automated framework can distinguish the more tough cases.

Unlike lexical compounds where there is no/little knowledge of how the compounding took place and the compounds became popular, there are sociological influences in the formation of hashtag compounds. Hashtag compounds can be made popular artificially. For example, \#AmazonPrimeDay, \#WWW2015, \#KDD2015 etc. There are also spontaneous pressures of hashtag compound formation. These kind of compounds are generally conversational hashtags or Idioms. Idioms actually have different spreading mechanism as shown in~\cite{romero}. They have high stickiness and low persistence. In table~\ref{spon}, we further show the differences of these two kinds of hashtag compounds in terms of some of their statistical properties. The first four compounds are examples of Idioms ( i.e., spontaneous formation of hashtags) whereas the remaining four are examples of forced/influenced hashtag compounds. We observe that in general, the spontaneous compounds have lesser number of mentions per tweets and lesser no. of collocations with 
other hashtags compared to the forced/influenced compounds. 
Also, the forced hashtag compounds spread via multiple mentions in early stage of propagation unlike the spontaneous ones.
\begin{table}[h]
\small
\centering
\caption{Differences in spontaneous (first four rows) and forced/influenced (last four rows) hashtag compounds}
  \label{spon}
  \resizebox{8.5cm}{!}{
 \begin{tabular}{|p{4cm}|p{1cm}|p{1cm}|p{1cm}|p{1cm}|}
\hline
Hashtag compounds & no.of mentions per tweet & no. of retweets per tweet & no. of collocations per tweet & no. of mentions in first 50 tweets \\\hline
\#TheBestFeelingInARelationship & 0.074 & 0.370 & 0.148 & 2 \\\hline
\#10WorstFeelings & 0.041 & 0.434 & 0.097 & 0\\\hline
\#YouKnowItsRealWhen & 0.071 & 0.372 & 0.208 &3 \\\hline
\#RelationshipTips & 0.023 & 0.498 & 0.175 &1 \\\hline
\#CMTAwards & 0.446 & 0.225 & 0.401 & 12\\\hline
\#JessicaForTheWin & 0.324 & 0.294 & 0.853 & 11\\\hline
\#SmartGalaxyS3 & 0.495 & 0.430 & 0.183 & 16\\\hline
\#BringBackToonami & 0.565 & 0.214 & 0.159 & 20\\\hline
  \end{tabular}
  }
\end{table}

\section{Conclusions and future works}
In this paper, we investigated various socio-linguistic properties responsible for hashtag compound formation and proposed a model to early predict popular hashtag compounds. To the best of our knowledge, this is the first study which deals with hashtag compounding and its adoption at a large scale over a very popular social media.

Our proposed prediction framework achieves a high accuracy of 77.07\% with a high precision and recall. We observe that the tweet content features are most discriminative compared to others. Among the tweet content features, the overlap features like n-gram overlap and word overlap are the most significant ones. The baseline accuracy based on human judgment experiment is only 48.7\%. This indicates that humans are not able to predict popular compound formation efficiently; in contrast, our model suitably informed with the right set of discriminative features is able to predict the popular compounds highly accurately with an overall $\sim$58\% improvement on the baseline. We also perform long term predictions after $T$ = 6 and 10 months after compounding and achieve 77.5\% and 79.13\% accuracy respectively. Correspondence analysis of the results obtained from the human judgments and the automatic framework shows that while the former is able to distinguish between the relatively easier cases, the latter is 
more successful in classifying the harder cases.

There are quite a few other interesting directions that can be explored in future. One such direction could be to study the lexical compounding on large scale data available in the form of millions of digitized books and newspaper archives. This study, we believe, can have important contributions to many NLP applications.
\section{Acknowledgments}
The authors thank the anonymous reviewers whose suggestions greatly helped improving the paper. The authors also thank Prof. Chris Biemann, TU Darmstadt for providing them with a historical Twitter 1\%
random sample data. This work has been supported by Microsoft Corporation and Microsoft Research India under the
Microsoft India PhD fellowship award.
\bibliographystyle{abbrv}
\bibliographystyle{SIGCHI-Reference-Format}
\bibliography{ref}

\end{document}